\documentclass[pre,aps,floatfix,a4paper,preprint]{revtex4-1}
\pdfoutput=1

\usepackage{graphicx}
\usepackage{hyperref}
\usepackage{amsmath}
\usepackage{float}
\begin{document}

\title{Influence of Rough and Smooth Walls on Macroscale Granular Segregation 
Patterns}

\author{Umberto D'Ortona}
\email{umberto@l3m.univ-mrs.fr}
\author{Nathalie Thomas} 
\affiliation{Aix-Marseille Universit\'e, CNRS, Centrale Marseille, M2P2 UMR 7340, 13451, Marseille, France}
\author{Richard M. Lueptow}
\affiliation{Department of Mechanical Engineering, 
         Northwestern University, Evanston, Illinois 60208, USA}

\date{\today}
\begin{abstract}
Size bidisperse granular materials in a spherical tumbler segregate into
two different patterns of three bands with either small particles at the
equator and large particles at the poles or vice versa, depending upon the
fill level in the tumbler. Here we use discrete element method (DEM)
simulations with supporting qualitative experiments to explore the effect
of the tumbler wall roughness on the segregation pattern, modeling the
tumbler walls as either a closely packed monolayer of fixed particles
resulting in a rough wall, or as a geometrically smooth wall. Even though
the tumbler wall is in contact with the flowing layer only at its
periphery, the impact of wall roughness is profound.  Smooth walls tend
toward a small-large-small (SLS) band pattern at the pole-equator-pole at
all but the highest fill fractions; rough walls tend toward a
large-small-large (LSL) band pattern at all but the lowest fill fractions.
This comes about because smooth walls induce poleward axial drift of small
particles and an equator-directed drift for large particles, resulting in
an SLS band pattern.  On the other hand, rough walls result in both sizes
of particles moving poleward at the surface of the flow, but due to radial
segregation, small particles percolate lower in the flowing layer where
there is a return drift toward the equator while large particles remain at
the surface near the pole, resulting in an LSL band
pattern.  The tendency toward either of the two band patterns depends
on the fill level in the tumbler and the roughness of the
tumbler's bounding wall.
\end{abstract}

\maketitle
\section{Introduction}

Discrete element method (DEM) simulations are used extensively to study 
the flow and segregation of granular materials in many situations as a 
predictive tool and access to 
data that are otherwise difficult to obtain experimentally. 
One of the key aspects of any simulation approach is the implementation of boundary 
conditions at walls. Two types of wall boundary conditions 
can be implemented in DEM simulations:
1) geometrically smooth surfaces \cite{MoakherShinbrot00,ChenLueptow11,TaberletLosert04,Rapaport02,TaberletNewey06}, which are assumed
 to have infinite mass and a specified radius of curvature (infinite for 
planar walls); and 2) a geometrically 
rough surface made up of a closely packed monolayer of fixed particles 
conforming to the geometry of the wall surface (for example, 
see~\cite{DaCruzEmam05,McCarthyOttino98,PoschelBuchholtz95,JuarezChen11,BertrandLeclaire05}). However, a recent study of monodisperse 
flow in a spherical tumbler suggests that the results using the latter 
approach, often called a “rough wall,”
differ from those using a smooth wall, not only locally at the particle scale
but also globally across the entire flowing layer \cite{DOrtonaThomas15}.

In this paper, we explore the impact of rough and smooth walls
on the axial segregation of bidisperse 
particles in a partially-filled spherical tumbler rotating with angular 
velocity $\omega$ about a horizontal axis.
We consider the situation where the free surface is essentially 
flat and continuously flowing. In this regime, the surface of the flowing 
layer maintains a dynamic angle of repose $\beta$  with respect to horizontal
that depends on the frictional properties, the diameter $d$ of the particles,
and the rotational speed of the tumbler 
\cite{PignatelLueptow12,TaberletRichard03,MeierLueptow07,duPontGondret03}. 
In experiments with spherical tumblers approximately half filled with 
a 50\%-50\% size bidisperse mixture of particles and with smooth walls, large 
particles accumulate near the poles of the tumbler with a band of small 
particles at the equator \cite{GilchristOttino03,ChenLueptow09}.  This pattern is 
inverted for lower fill fractions so that small particles accumulate near 
the poles with a band of large particles at the equator \cite{ChenLueptow09}. 
We note that multiple
 bands of small and large particles occur for bidisperse mixtures in long 
rotating cylindrical tumblers, which are used in applications for materials  ranging from 
foodstuffs to mining to cement, typically after O(10) to O(100) 
rotations and having a wavelength of about one tumbler diameter 
\cite{JuarezLueptow08} under a wide range of conditions 
\cite{JuarezLueptow08,Oyama39,DonaldRoseman62,Nakagawa94,ZikLevine94,HillKakalios94,HillKakalios95,JainLueptow01}. In the cylindrical
 tumbler case, however, large particles segregate near the flat end walls of the 
tumbler \cite{JuarezLueptow08,HillKakalios94,FiedorOttino03}, as a consequence of radial 
segregation combined with the nonuniform axial distribution of velocity in 
the flowing layer due to friction at the endwall \cite{ChenLueptow10}.
This mechanism cannot occur in the spherical tumblers studied here.

\begin{figure}[btph]
\includegraphics[width=0.45\linewidth]{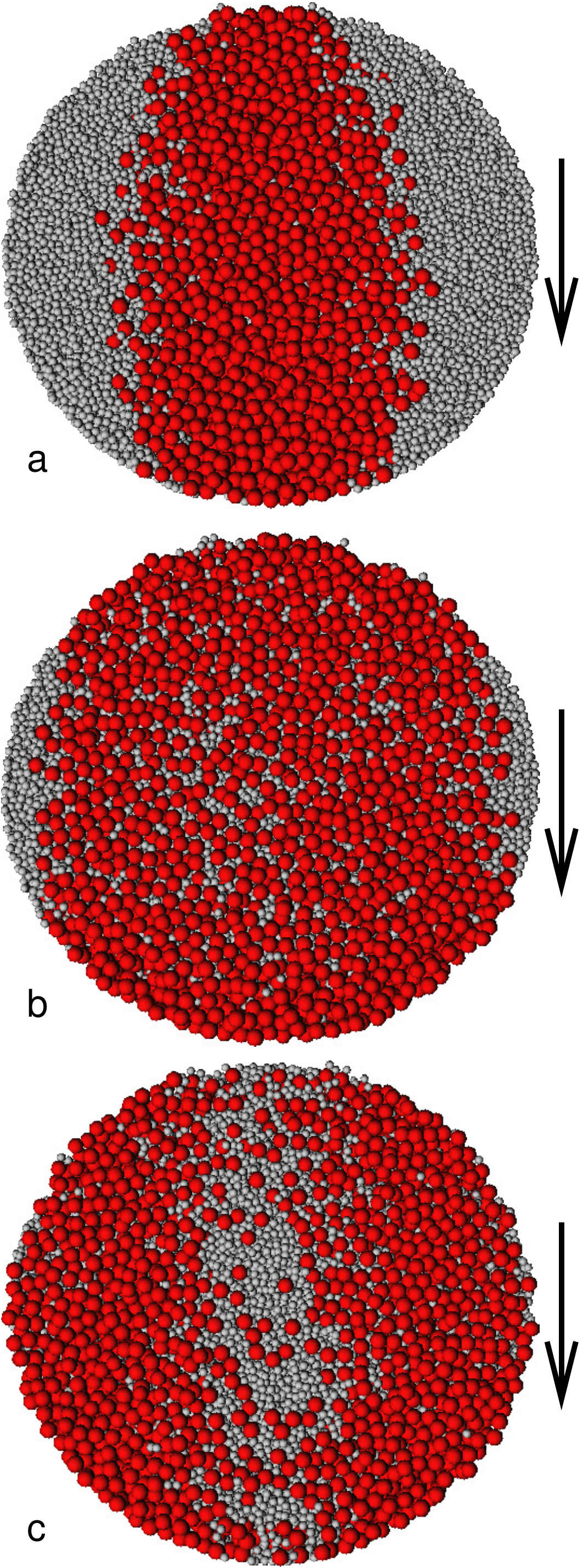}
\caption{(Color online) Steady-state surface segregation patterns for equal volumes of 2~mm and 4~mm 
particles in a
30\% full 14~cm diameter spherical tumbler rotated at 15~rpm for $t =$~200~s. 
The tumbler walls are (a) smooth, (b) rough wall of 1.5~mm particles and (c)
rough wall of 2~mm particles. The rotation axis is horizontal and arrows show
the direction of the surface flow.}
\label{roughsmooth30p}
\end{figure}

Although DEM simulations of spherical tumblers with smooth walls readily 
reproduce the segregation experiments in an acrylic spherical tumbler 
\cite{ChenLueptow09}, the mechanism for the inversion of the segregation bands from 
large-small-large (LSL) for higher fill fractions to small-large-small (SLS) 
at lower fill fractions remains unresolved. 
 In an initial effort to understand the segregation mechanism, we attempted 
further DEM simulations at different fill fractions in spherical tumblers. 
 However, to simplify the implementation of the simulations, we performed the
simulations with a rough wall boundary conditions. The results are
dramatically different from the simulations with smooth walls, as shown in 
Fig. \ref{roughsmooth30p}.  For otherwise identical systems (same particle 
sizes, rotational speed, tumbler diameter, and fill fraction), the surface
segregation 
pattern changes from SLS for smooth walls (Fig. \ref{roughsmooth30p}(a)) to LSL 
for rough walls made of 2~mm diameter particles
(Fig.~\ref{roughsmooth30p}(c)). Using an intermediate wall particle 
size of 1.5~mm results in no significant surface band 
formation at all (Fig. \ref{roughsmooth30p}(b)). 

The unexpectedly strong
influence of wall roughness on band formation led us to first study
monodisperse flows in a spherical tumbler \cite{DOrtonaThomas15}. For 
monodisperse flows, the wall
roughness strongly affects the particles trajectories, even far from walls.
Particle trajectories at the free surface 
curve further toward the poles for smooth walls than for rough walls. 
However, the particle trajectories curve back toward the tumbler equator more 
in the smooth
case as well, resulting in a smaller net poleward drift at the surface
 for smooth walls than
for rough walls.

In this paper, we examine through both DEM simulations and 
qualitative experiments 
the impact of wall boundary roughness on band formation in bidisperse flows.

\section{Qualitative Experiments}

\begin{figure}[btph]
\includegraphics[width=0.65\linewidth]{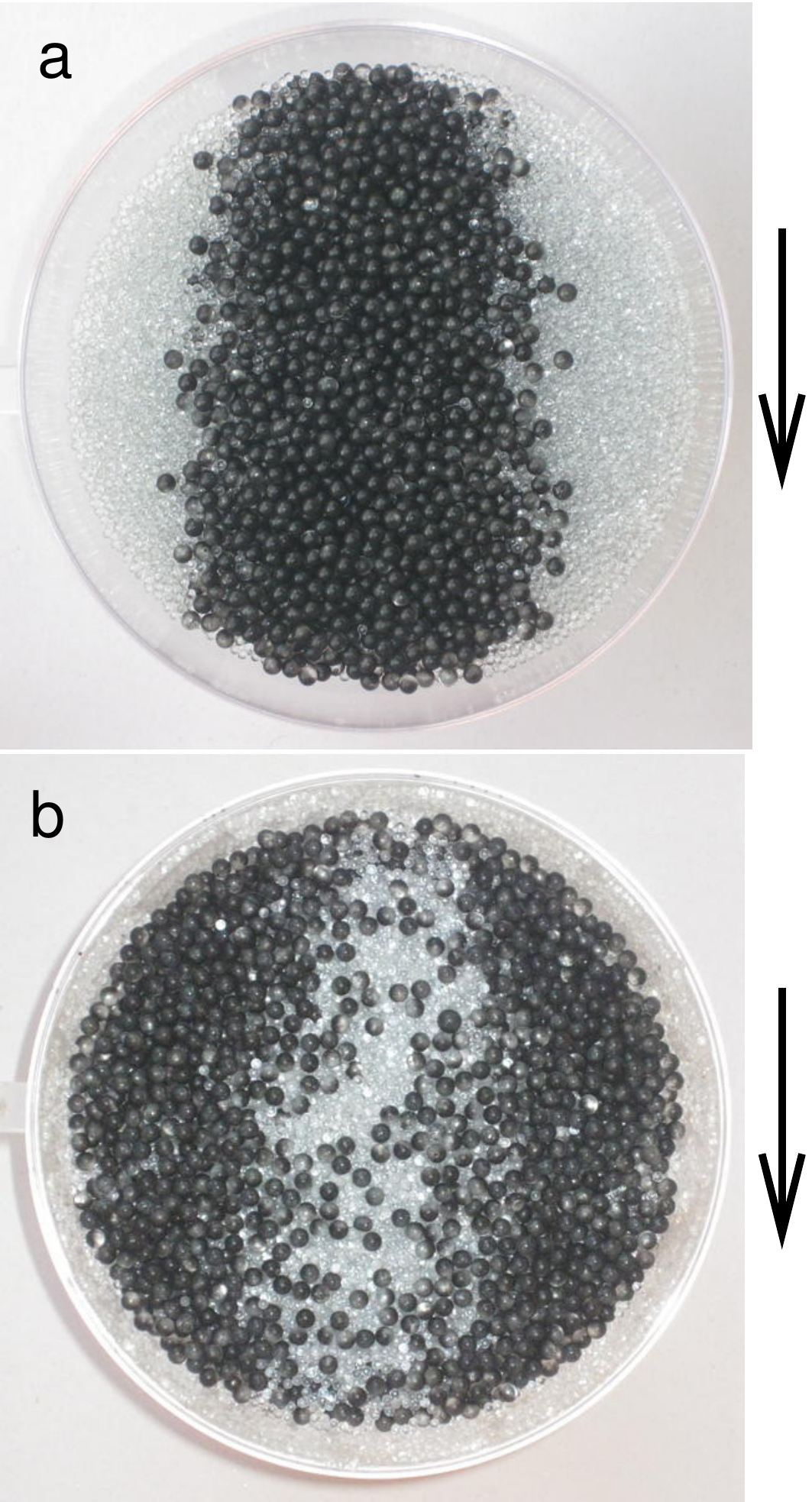}
\caption{Surface segregation patterns for equal volumes of 2~mm (transparent) and 4~mm (black) glass particles in a 30\% full spherical acrylic tumbler 
rotated at 15 rpm for 400~s (100 rotations).
The tumbler walls are (a) smooth, (b) rough with 2~mm particles. The rotation
axis is horizontal and arrows show the direction of the surface flow.}
\label{exproughsmooth}
\end{figure}

Since the different results for smooth and rough walls were initially
 obtained via DEM simulations, it was crucial
to confirm that the predicted segregation patterns did in fact occur
experimentally. Qualitative experiments were performed using clear acrylic
spheres consisting of two mating hemispheres of diameter $D=14~$cm rotated
by an electric motor at 14.7~rpm about a horizontal axis.
The tumbler was filled to 30\% by volume with equal volumes
of $d = 2$~mm and $d = 4$~mm diameter spherical 
glass particles.
For the rough wall case, 
the small particles were bonded to
the wall of a tumbler using epoxy, 
thereby reducing the tumbler
inner diameter to 13.6~cm. The tumbler was rotated for approximately 100
rotations and stopped so that all the particles were in one hemisphere.
Then the upper hemisphere of the spherical tumbler was removed to obtain an image of the surface
segregation pattern. The SLS 
segregation pattern
occurs for the smooth tumbler wall (Fig.~\ref{exproughsmooth}(a)), while the LSL segregation pattern
occurs for the rough wall having 2~mm particles bonded to it (Fig.~\ref{exproughsmooth}(b)), confirming
the validity of the DEM results and the surprising effect of the wall roughness
on the segregation pattern.

\section{DEM Simulations}

For the
DEM simulations, a standard linear-spring and viscous damper force model
\cite{ChenLueptow08,SchaferDippel96,Ristow00,CundallStrack79} was used to calculate
the normal force between two contacting particles: 
$F_n^{ij}=[k_n\delta - 2 \gamma_n m_{\rm eff} (V_{ij} \cdot \hat r_{ij})]\hat r_{ij}$, 
where $\delta$ and $V_{ij}$ are the particle overlap and the relative
velocity $(V_i - V_j)$ of contacting particles $i$ and $j$ respectively; 
$\hat r_{ij}$ is
the unit vector in the direction between particles $i$ and $j$; 
$m_{\rm eff} = m_i m_j/(m_i + m_j)$ is the reduced mass of the two particles; 
$k_n = m_{\rm eff} [( \pi/\Delta t )^2 + \gamma^2_n]$ is the normal stiffness 
and $\gamma_n = \ln e/\Delta t$ is the normal damping, where $\Delta t$
is the collision time and $e$ is the restitution coefficient \cite{ChenLueptow08,Ristow00}. A
standard tangential force model \cite{SchaferDippel96,CundallStrack79} with
elasticity was implemented: $F^t_{ij}= -\min(|\mu F^n_{ij}|,|k_s\zeta|){\rm sgn}(V^s_{ij})$, where
$V^s_{ij}$ is the relative tangential velocity of two particles \cite{Rapaport02},
$k_s$ is the tangential stiffness, $\mu$ the Coulomb friction coefficient and $\zeta(t) = \int^t_{t_0} V^s_{ij} (t') dt'$ is the net
tangential displacement after contact is first established at time $t = t_0$.
The velocity-Verlet algorithm \cite{Ristow00,AllenTildesley02} was used to 
update
the position, orientation, and linear and angular velocity of each
particle. Tumbler walls were modeled as both smooth surfaces (smooth walls)
and as a monolayer of bonded particles (rough walls). Both wall conditions
had infinite mass for calculation of the collision force between the
tumbling particles and the wall. 

The spherical tumbler of radius $R=7$~cm was filled to volume fraction $f$ with equal volumes of
small and large particles of diameter 2 and 4~mm; gravitational
acceleration was $g$ = 9.81~m~s$^{-2}$; particle properties correspond to cellulose
acetate: density $\rho =$ 1308~kg~m$^{-3}$, restitution coefficient $e = 0.87$ \cite{DrakeShreve86,FoersterLouge94,SchaferDippel96}. The two species were initially randomly
distributed in the tumbler with a total of about $5 \times 10^4$ particles in a
typical simulation. To avoid a close-packed structure, the particles had a
uniform size distribution ranging from 0.95$d$ to 1.05$d$. Unless otherwise 
indicated, the friction
coefficients between particles and between particles and walls was set to 
$\mu = 0.7$.  The collision time was $\Delta t$ =10$^{-4}$ s, consistent with previous
simulations \cite{TaberletNewey06,ChenLueptow11,ZamanDOrtona13} and sufficient for
modeling hard spheres \cite{Ristow00,Campbell02,SilbertGrest07}.  These
parameters correspond to a stiffness coefficient $k_n = 7.32\times 10^4$ (N m$^{-1}$)
\cite{SchaferDippel96} and a damping coefficient $\gamma_n = 0.206 $~kg~s$^{-1}$.  The
integration time step was $\Delta t/50 = 2\times 10^{-6}$~s to meet the requirement of
numerical stability \cite{Ristow00}.

\section{Results}
\subsection{Segregation patterns}

\begin{figure}[btph]
\includegraphics[width=0.95\linewidth]{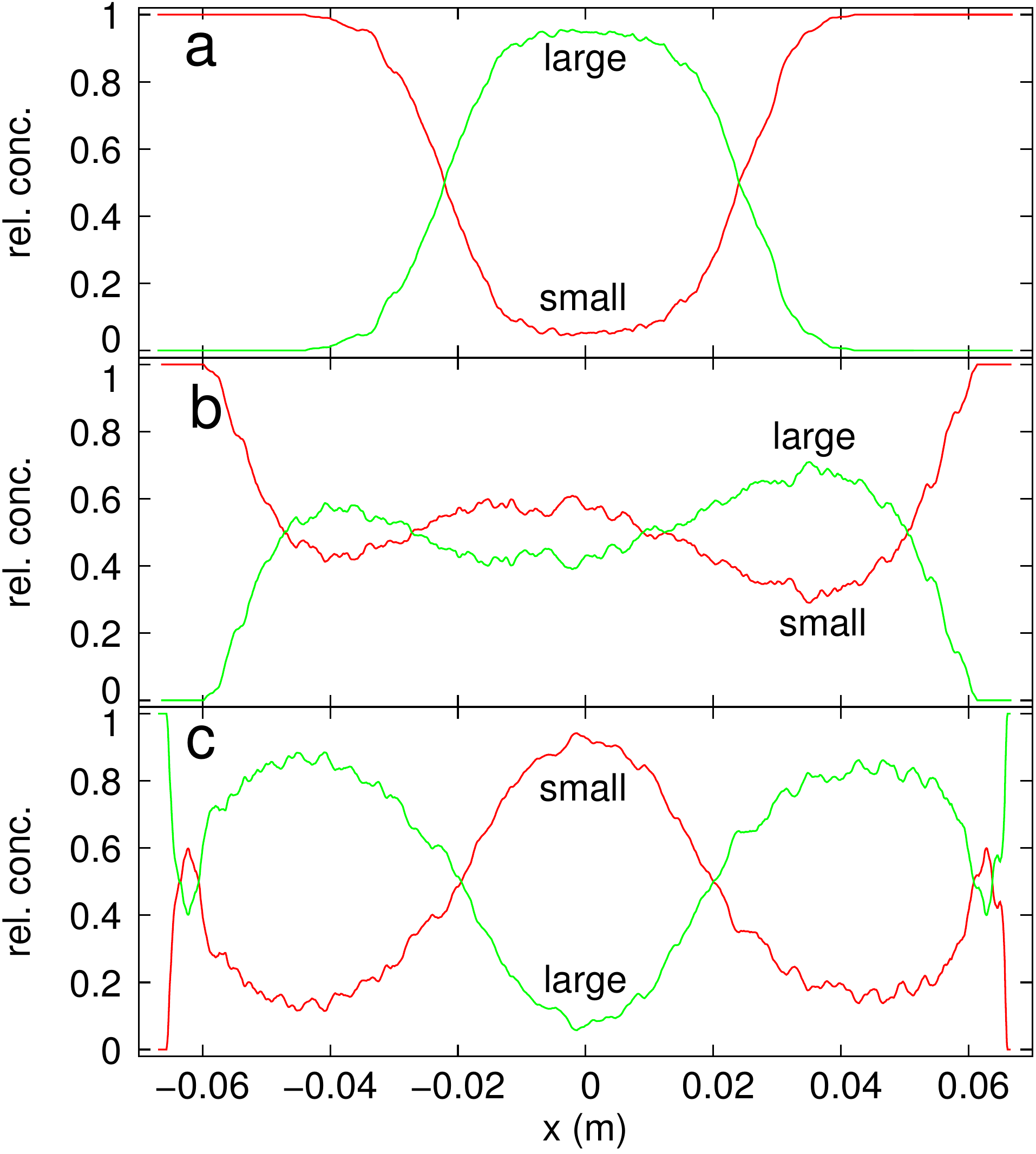}
\caption{(Color online) Concentration profiles of small and large particles 
corresponding to Fig.~\ref{roughsmooth30p}: (a) smooth, (b)
rough 1.5~mm particles and (c) rough 2~mm particles.}
\label{proroughsmooth}
\end{figure}

\begin{figure}[btph]
\includegraphics[width=0.65\linewidth]{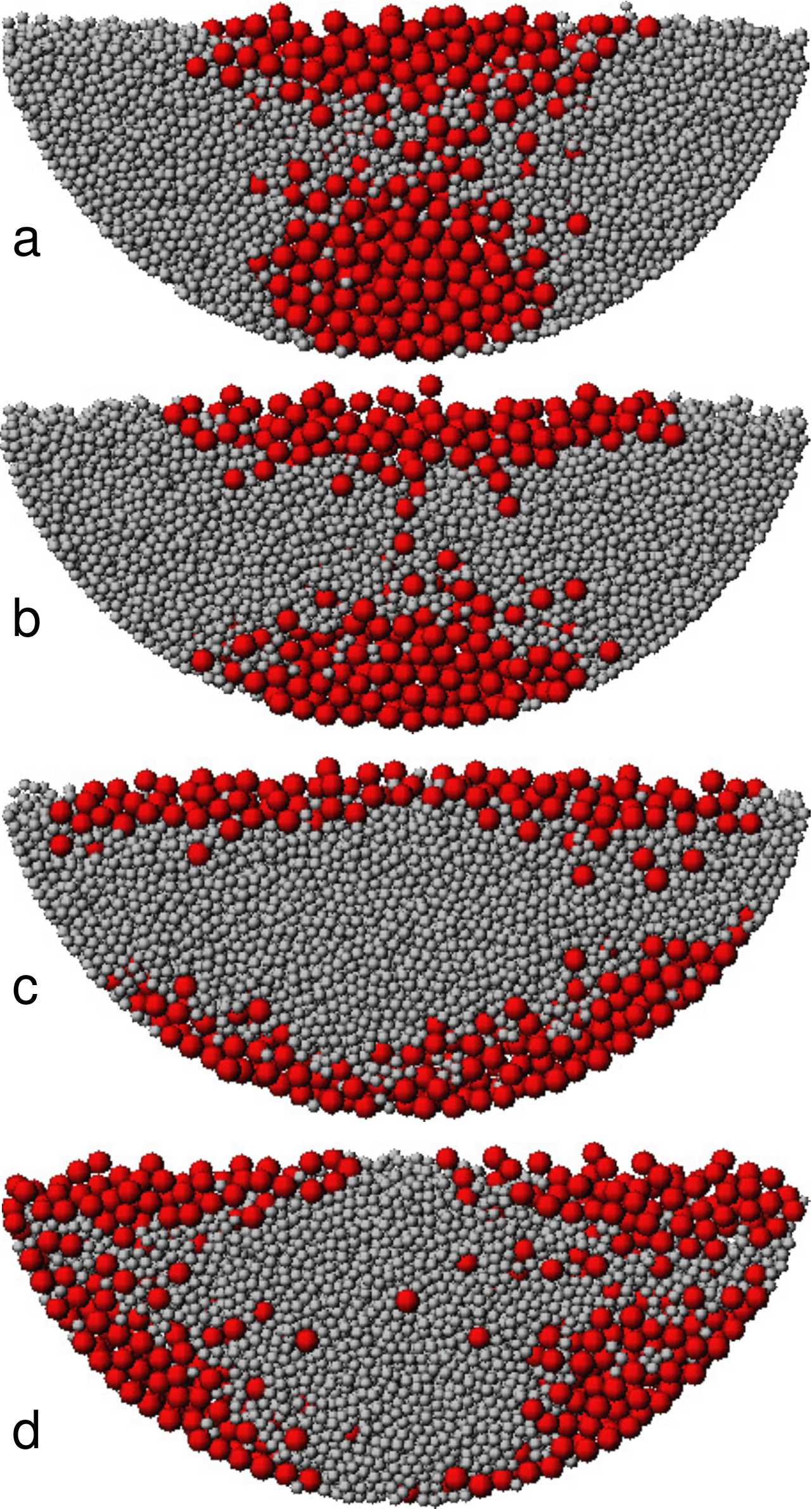}
\caption{(Color online)
Cross-section of the segregation patterns in a vertical plane 
containing the axis of rotation corresponding to the surface 
patterns in Fig.~\ref{roughsmooth30p} and concentration profiles in 
Fig.~\ref{proroughsmooth}. The tumbler walls are (a) smooth, (b) rough 
with 1~mm particles, (c) rough with 1.5~mm particles and (d) rough 
with 2~mm particles.}
\label{sliceroughsmooth30p}
\end{figure}

The steady-state concentration profiles corresponding to the segregation 
patterns in Fig.~\ref{roughsmooth30p} are shown in Fig.~\ref{proroughsmooth}. 
A relative concentration of 1.0 corresponds to pure particles of one size.
The profiles were obtained by determining the size of the particles 
intersecting a series of planes perpendicular to the axis of rotation (with 
$x=0$ at 
the equator of the tumbler) and extracting the volume concentration for each 
species in that plane.  This approach allows a much higher axial resolution 
for the concentration measurements than can be achieved with simple binning.
For the SLS pattern in the smooth wall tumbler (Fig.~\ref{proroughsmooth}(a)), the three bands 
are nearly pure (one particle size or the other for most of the width of each
band).  On the other hand, for the LSL pattern in the 2~mm rough wall 
tumbler (Fig.~\ref{proroughsmooth}(c)), the bands are less pure and the 
transition between 
bands is not as sharp. The 1.5~mm rough wall (Fig.~\ref{proroughsmooth}(b))
results in a configuration intermediate between the two other cases. 
The underlying reason for the nature of the 
concentration profiles is evident when viewing a cross-section in a vertical 
plane and containing the axis of rotation 
(Fig.~\ref{sliceroughsmooth30p}). 
For the SLS segregation pattern, the 
small particles and large particles form more distinct bands through the depth 
of the particle bed (Fig. \ref{sliceroughsmooth30p}(a)), while for the LSL 
surface pattern, the bands are much less sharp (Fig. \ref{sliceroughsmooth30p}(d)).  
More interesting are the 1.0 and 1.5~mm rough wall cases 
(Fig. \ref{sliceroughsmooth30p}(b) and \ref{sliceroughsmooth30p}(c)). 
In both cases, there is a core of small particles across the entire width 
of the bed of particles with large particles surrounding the core.
Fig. \ref{sliceroughsmooth30p}(b) for 1~mm rough walls shows a configuration
 where the surface pattern is still SLS, but the band of large particles is 
less pure.  In Fig~\ref{sliceroughsmooth30p}(c), particles have segregated 
radially but not axially, except right at the poles leading to no visible 
surface bands.

\subsection{Axial segregation index}
The segregation patterns develop over time. A convenient means to 
quantify the degree of segregation and its time evolution is an 
axial segregation index, defined as:  
\begin{equation}
I= \frac{1}{R}\left(\frac{\displaystyle\sum_{i={\rm large}} |x_i|v_i }{\displaystyle \sum_{ i={\rm large}} v_i}
- \frac{\displaystyle\sum_{ i={\rm small}} |x_i|v_i }{\displaystyle \sum_{ i={\rm small}} v_i}\right)
\end{equation}
where $x_i$ is the position along the rotation axis with the equator at $x=0$, $v_i$
is the volume of the particle species $i$, $R$ is the radius of the sphere, and the 
summations are over the large and small particles.  
The axial segregation index is positive for LSL and negative for SLS.  The 
limit for perfect segregation is always less than one, but depends on the 
fill fraction.  For a 50\% full tumbler, perfect segregation would result 
in a segregation index of approximately $\pm$0.4 based on the tumbler's 
spherical shape.

\begin{figure}[htbp]
\includegraphics[width=0.8\linewidth]{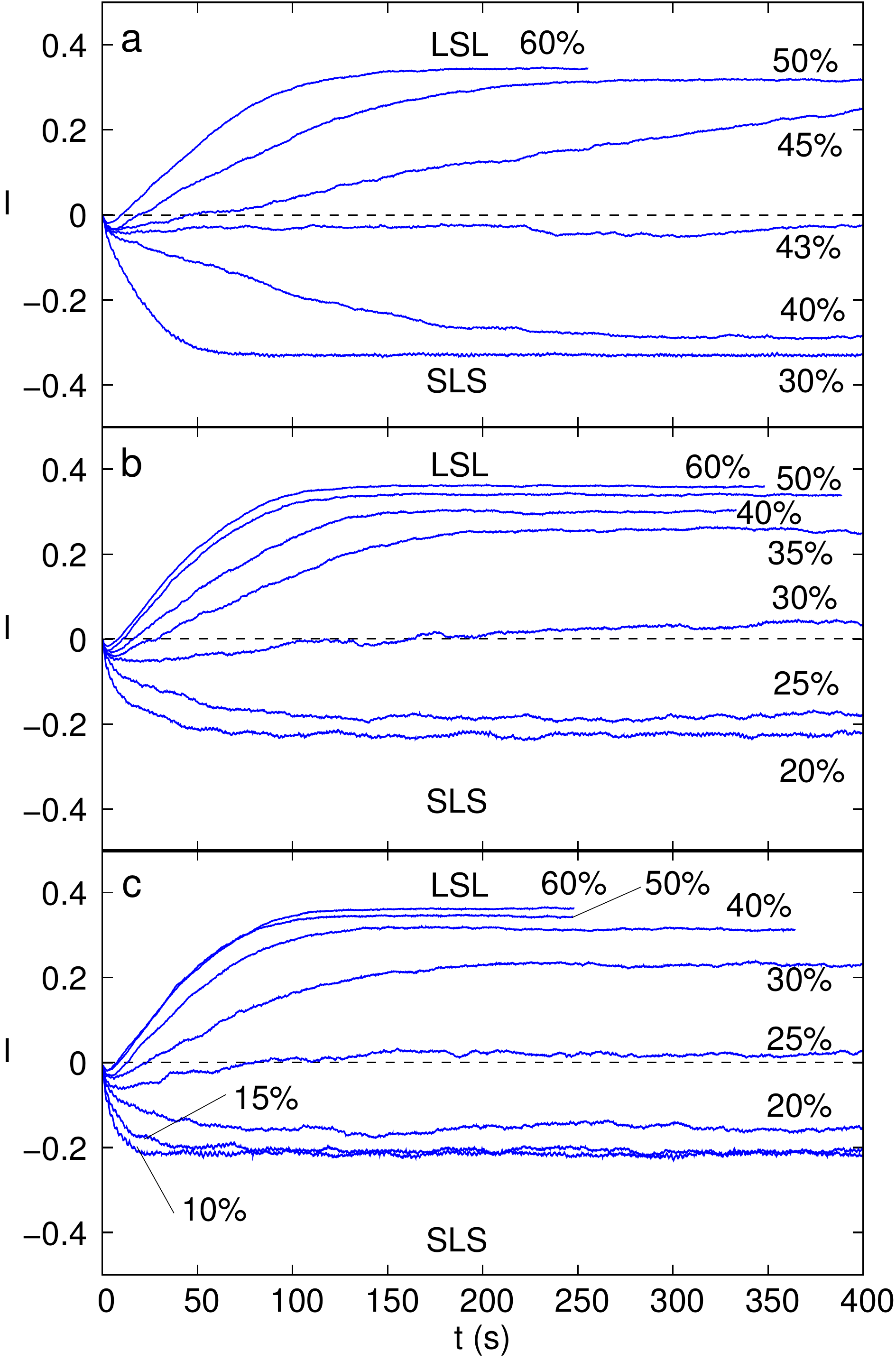}
\caption{Evolution of the axial segregation index $I$ for equal volumes of 2~mm
and 4~mm particles rotated at 15~rpm for several tumbler fill 
fractions (volume percentages indicated in the figure) and wall roughnesses: 
(a) smooth, (b) rough with 1.5~mm particles and (c) rough with 2~mm particles.}
\label{ios0-2mm}
\end{figure}
Figure \ref{ios0-2mm} shows the time evolution of the axial segregation index 
for the three wall roughness cases
in Fig.~\ref{roughsmooth30p} (with increasing wall roughness from top to bottom) for a range
of fill fractions. Note that simulations were conducted until the segregation
index reached its asymtotic value. Only the time evolution up to 400~s is presented here, but
in some cases, simulations reached 1000~s. Regardless of the wall roughness, the
segregation pattern is usually LSL ($I> 0$) for larger fill fractions and SLS 
($I< 0$)
for smaller fill fractions, consistent with experimental results for smooth
walls \cite{ChenLueptow09}. Further note that LSL segregation ($I> 0$) is achieved
more slowly than SLS segregation ($I < 0$) for a smooth wall, also consistent
with experimental results \cite{ChenLueptow09}. However, this is to be expected because the
time that particles spend in the flowing layer depends directly on the fill
fraction. For low fill fractions, particles pass through the flowing layer
more frequently than at high fill fractions for the same elapsed time 
\cite{ZamanDOrtona13}. Since only particles in the flowing layer have the opportunity
to rearrange themselves (unlike particles below the flowing layer, which are
locked into place in the bed of particles in solid body rotation), one could
reasonably expect the segregation pattern to appear more quickly for low fill
fractions than for high fill fractions. Similar results occur for rough
walls.

\begin{figure}[htbp]
\includegraphics[width=\linewidth]{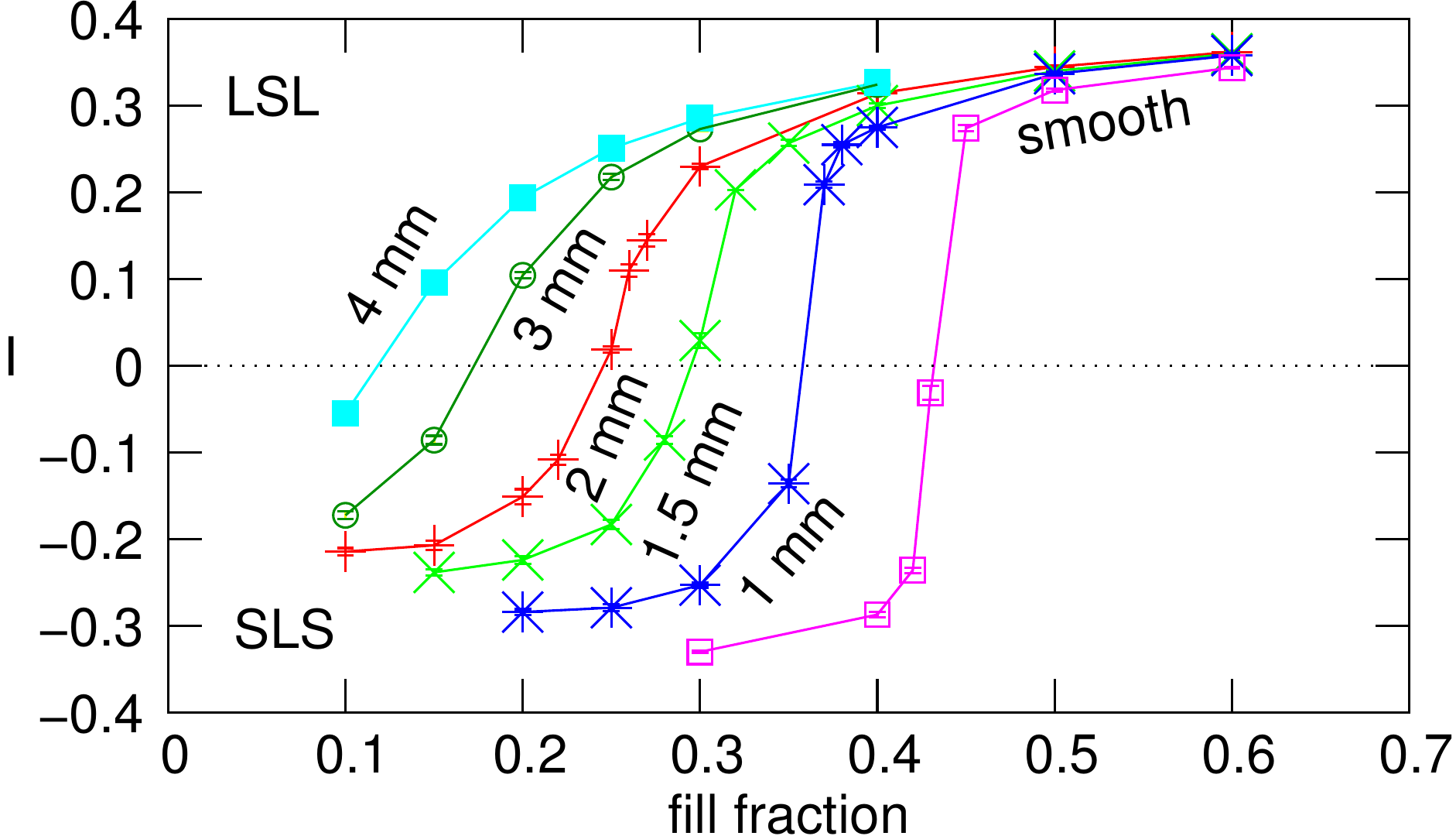}
\caption{(Color online) Asymptotic value of the axial segregation index as 
function of fill fraction for several wall roughnesses for equal volumes of 
2~mm and 4~mm particles rotated at 15~rpm.  Error bars
(smaller than the symbols) are the standard deviation of the axial segregation index.}
\label{charnin}
\end{figure}

The transition between 
LSL and SLS occurs at different fill fractions depending on the wall
roughness (Fig.~\ref{charnin}).
The
fill fraction for transition decreases from 
43\% for the smooth wall case to 12\% for the 4~mm rough wall case.  At
high fill fractions, the segregation 
gets close to nearly perfect LSL segregation, regardless of the wall roughness. 
Increasing the wall roughness favors LSL segregation. 
At low fill fractions, 
the degree of SLS segregation depends on the roughness of the wall, with the
smooth wall having the greatest segregation. However the magnitude of the
segregation index for SLS segregation is not as large as with LSL segregation,
except for smooth walls.
Note that fill fractions lower than those shown in Fig.~\ref{charnin} for
roughnesses of 2~mm or less result 
in slip of the particle bed with respect to the tumbler, so they are not
included in the figure.
The transition from SLS to LSL at $I=0$ is steepest for smooth walls.

\begin{figure}[btph]
\includegraphics[width=0.8\linewidth]{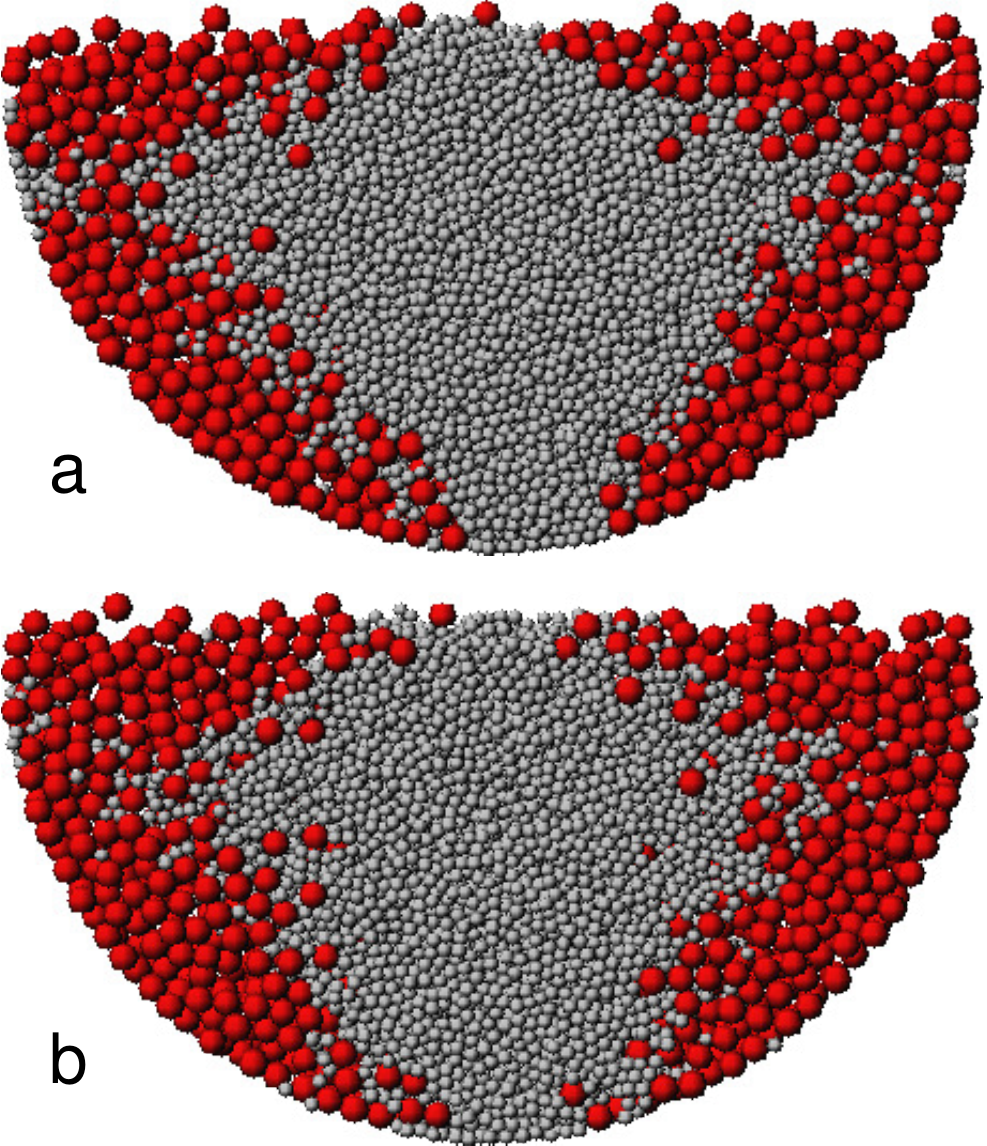}
\caption{(Color online)
Cross-section of the segregation patterns in a vertical plane 
containing the axis of rotation corresponding to a sphere filled
at 50\% with equal volumes of 2~mm and 4~mm particles. The tumbler walls are (a)
smooth and (b) rough with 2~mm particles.}
\label{bidi50p}
\end{figure}

It is helpful to consider the axial segregation index in Fig.~\ref{charnin} in 
the context of the segregation patterns in Fig.~\ref{sliceroughsmooth30p}.  
Moving upward along a vertical line in Fig.~\ref{charnin} at a fill fraction 
of 30\% starts with strong SLS segregation corresponding to 
Fig.~\ref{sliceroughsmooth30p}(a) for a smooth wall. Increasing the wall 
roughness to 1~mm particles corresponds to the development of a core of small 
particles at the equator, shown in Fig.~\ref{sliceroughsmooth30p}(b), reducing 
axial segregation index, though the poles retain nearly pure small particles. 
For a 1.5~mm rough wall the segregation index is approximately zero, 
corresponding to a core 
of small particles surrounded by large particles extending nearly to the poles,
shown in Fig.~\ref{sliceroughsmooth30p}(c). A rougher wall of 2~mm particles 
corresponds to strong LSL segregation in which large particles dominate near 
the poles, and small particles reach the flowing layer surface at the equator, 
shown in Fig.~\ref{sliceroughsmooth30p}(d).  This sequence is quite different 
from what 
occurs for a 50\% fill level.  There is almost no difference in the axial 
segregation index at a 50\% fill level in Fig.~\ref{charnin} for different wall 
roughness values.  The LSL segregation patterns shown in 
Fig.~\ref{bidi50p} at a 50\% fill volume for smooth and 2~mm rough walls bear 
out the similarity in the axial segregation index.

\subsection{Particle trajectories}
Previous results for the effect of wall roughness and fill level on particle 
trajectories in monodisperse flows provide some insights.
The trajectories in the flowing layer for both smooth and rough walls
 are curved \cite{ZamanDOrtona13,DOrtonaThomas15}.  This curvature is
negligible at the equator (at $x=0$~cm, which is a line of symmetry)
 and increases moving
toward the pole.
For particles near the surface of the flowing layer, the trajectory curvature 
for smooth walls is 
greater than that for
rough walls. However, the net poleward drift at the surface with each pass 
through the flowing layer for rough walls 
is larger than for smooth walls. Since
surface particles drift poleward, particles deeper in the flowing layer drift 
toward the equator to conserve mass. As the fill level is reduced, the curvature
of the particle trajectories increases for all roughnesses, and the poleward 
drift decreases only for smooth walls but does not change significantly for 
rough walls.

These monodisperse results can be used to explain the bidisperse segregation 
patterns. Two effects compete to select the segregation pattern.
First, particles are subject to a depth dependent poleward drift.
Since drift is larger at the surface \cite{DOrtonaThomas15}, large 
particles
that have segregated to the surface axially drift further poleward
than small particles below the surface.
Large particles remain at the surface due to radial 
segregation, so
they accumulate near the poles. Small particles below the surface in the
radially segregated core are transported by 
the global convection cells: from equator to pole nearer the 
surface and from pole to equator deeper in the flowing layer 
\cite{ZamanDOrtona13}.
The second effect, which counteracts the drift, is the high curvature of the 
particle trajectories at the surface of the flowing layer, which is
 typical of smooth 
walls \cite{DOrtonaThomas15}. It results in both small and large particles 
being carried further poleward in the upstream portion of the flowing layer.
Small particles tend to fall out of the flowing
layer sooner than larger particles due to percolation, thus depositing in the
fixed bed when they are closer to the poleward extrema of the trajectory than
large particles, which tend to stay near the surface to curve back more 
toward the equator to deposit at an
axial position near where they started. It is likely that for smooth walls,
this trajectory curvature
effect dominates, leading to the SLS pattern; for rough walls, where the 
trajectory curvature is smaller, the drift is very efficient and dominates 
\cite{DOrtonaThomas15} leading to the LSL pattern.

\begin{figure}[htbp]
\includegraphics[width=0.85\linewidth]{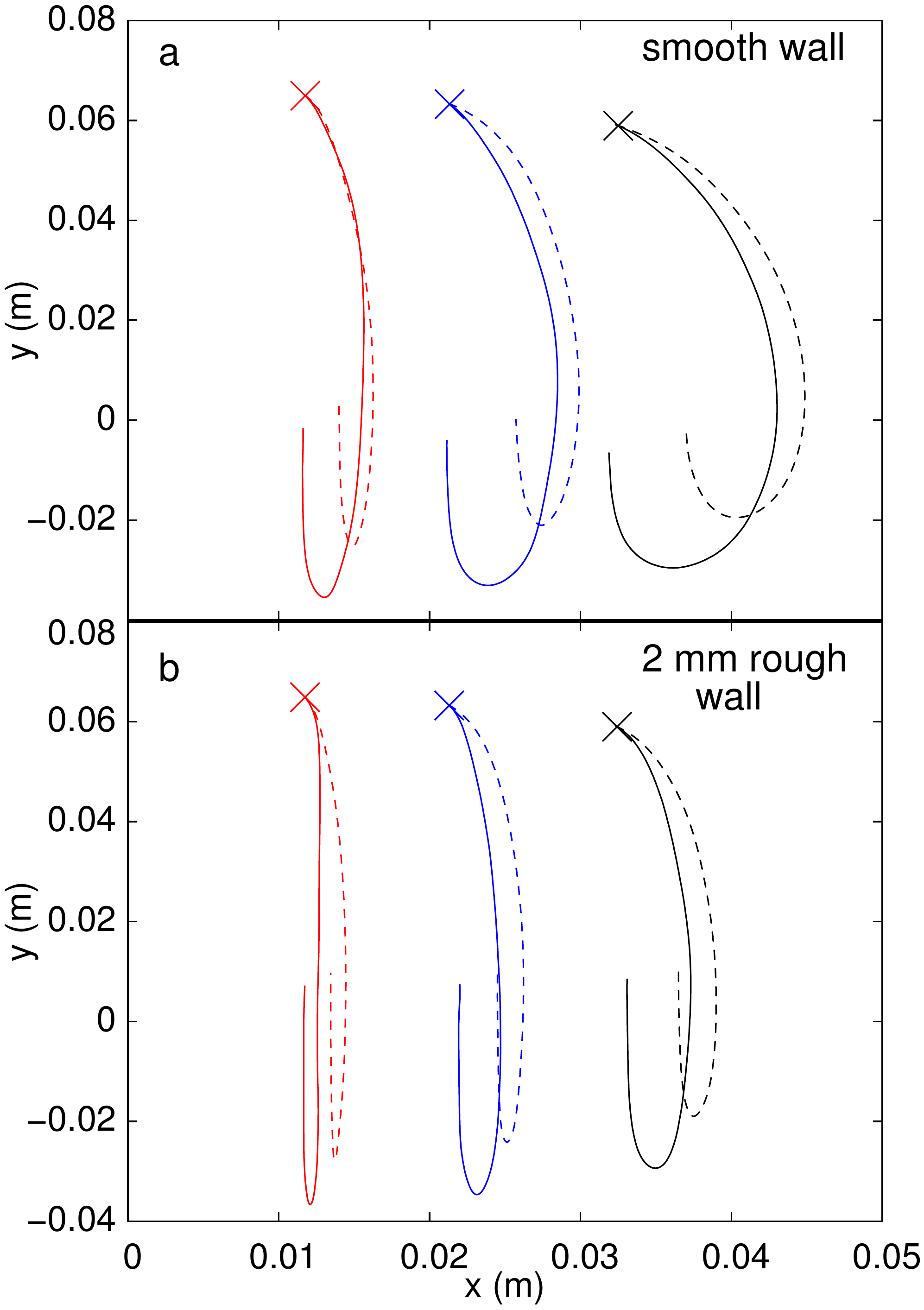}
\caption{(Color online) Comparison (top view) of pair of trajectories of 2~mm 
(dashed curves) and 4~mm (solid curves) particles starting from the same 
point (marked with an $\times$) in the flowing zone, at various $x$
coordinates, in a tumbler made of a) smooth walls or b) rough 2~mm walls
and filled at 30\%.}
\label{comparetraj30ptop}
\end{figure}

\begin{figure}[htbp]
\includegraphics[width=0.85\linewidth]{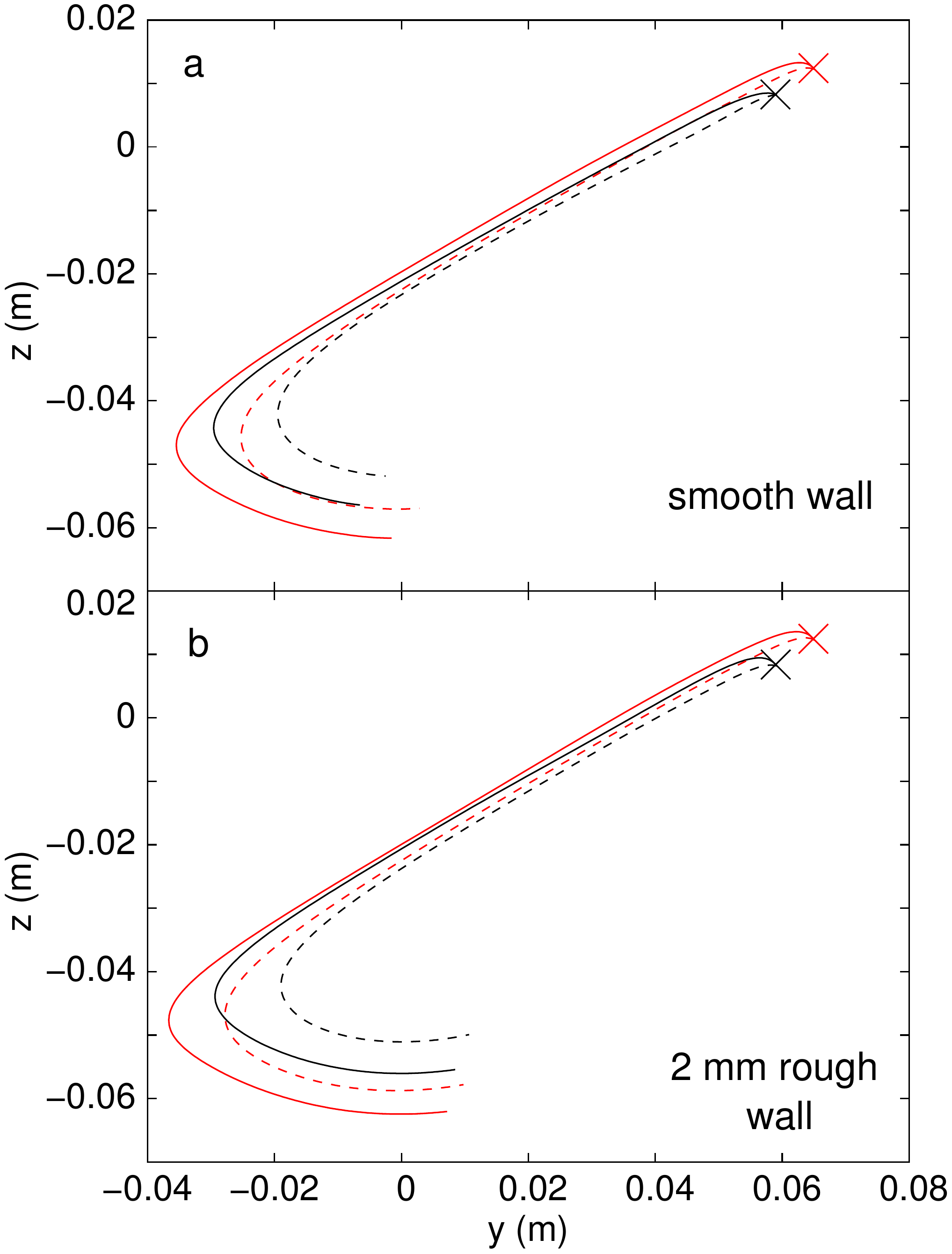}
\caption{(Color online) Comparison (side view) of pair of trajectories of 2~mm 
(dashed curves) and 4~mm (solid curves) particles starting from the same 
point (marked with an $\times$) in the flowing zone, but in various
$x$ positions, in a tumbler made of a) smooth walls or b) rough 2~mm walls
and filled at 30\%.}
\label{comparetraj30pside}
\end{figure}

To confirm this mechanism, we plot pairs of trajectories for the two species of 
particles (2 and 4~mm) starting from the same initial positions in the flowing
zone near the surface (specifically, when the center of a particle starts 
within a sphere of radius 1~mm that is 3~mm below the free surface), for tumblers with 
smooth or with 2~mm rough walls
in Figs.~\ref{comparetraj30ptop} and \ref{comparetraj30pside}. 
These trajectories are obtained by averaging thousands of individual particle
trajectories starting from the same initial coordinate during the first few 
seconds (between $t=2$~s and $t=6$~s) of the flow, before radial segregation is
achieved. Like the monodisperse case \cite{DOrtonaThomas15}, the trajectories 
in a tumbler with a smooth wall have a larger curvature 
 than those for the rough case 
(Fig.~\ref{comparetraj30ptop}). However, for the bidisperse case, large 
particles remain at surface while small particles sink deeper into the
flowing layer
(Fig.~\ref{comparetraj30pside}), regardless of whether the walls are 
smooth or rough.


The trajectory curvature results in the large particles tending
toward the equator and small particles toward the poles. This is most evident
comparing the trajectories starting at $x=0.032$~m, in the smooth wall tumbler 
(Fig.~\ref{comparetraj30ptop}(a)). This process dominates for 
smooth walls, leading to the SLS segregation pattern.
The process is quite rapid, and thus is always dominant in the first 
moments of the flow.  This is evident in Fig.~\ref{ios0-2mm} where the 
segregation indices in all cases are negative during the few first seconds, 
even in the cases of 
asymptotic LSL segregated systems. It is also evident that fully developed segregation
is reached more quickly for SLS than for LSL (also observed 
experimentally \cite{ChenLueptow08}).

On the other hand, for rough walls, the curvature of the trajectories is less, 
but the net drift of both particles is toward the pole. Due 
to the radial segregation, large particles stay at the surface, whereas
small particles are deeper in the flowing layer where there is 
a return flow toward the equator.
The drift effect tends toward an LSL segregation
pattern and dominates when the trajectory curvature is small.

Analogous results occur when  varying  the fill level. At low fill levels, 
particle trajectories are more curved \cite{DOrtonaThomas15}, which favors 
SLS segregation. At high 
fill levels, trajectories are nearly straight \cite{DOrtonaThomas15}, so 
the axial drift, 
as the particles segregate to different depths, dominates  
leading to the LSL segregation pattern.

\begin{figure}[htbp]
\includegraphics[width=0.60\linewidth]{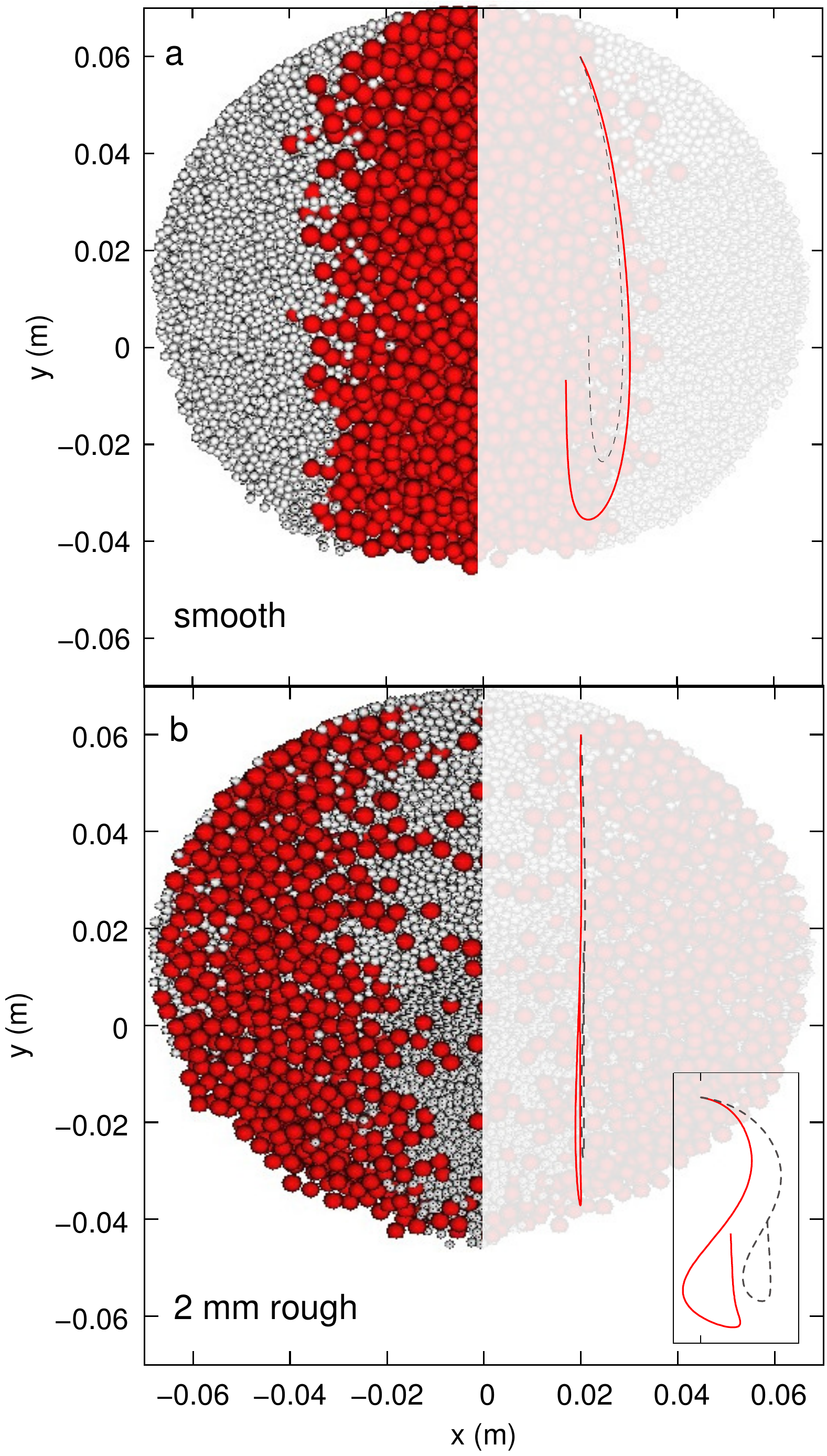}
\caption{(Color online) Steady state segregation pattern: 
comparison (top view) of a pair of trajectories of 
2~mm (dark grey dashed) and 4~mm
particles (red solid) starting from the same point in the flowing zone, at 
the border between the large and small particle regions,
in a tumbler made of a) smooth wall or b) rough 2mm walls and filled at 30\%. 
The surface of the flowing layer is at the angle of repose but viewed along
the gravity vector, so its perimeter is oval.
The inset in the bottom right of b) shows the same trajectories reduced
by a factor of 2 vertically and magnify by a factor of 10 horizontally.}
\label{trajbidicombi}
\end{figure}

At steady state for either SLS or LSL patterns, the particle trajectories appear
to stabilize the pattern. This is shown in Fig.~\ref{trajbidicombi} where
we compare the trajectories of large 
and small particles by plotting pairs of trajectories for the
two particle species at the boundary between the large and small particles.
In the case of smooth walls, the trajectories for the small and large
particles diverge substantially even after only one pass through the flowing 
layer, resulting in displacement toward the pole for small particles, 
and displacement toward the equator for large particles, reinforcing the SLS 
pattern.

This is not the case for the LSL segregation 
pattern for rough walls where displacements are very small, and both toward
the pole (see inset of Fig.~\ref{trajbidicombi}(b)).
Both species have nearly the same axial position after one pass through the 
flowing layer.
The axial segregation occurs indirectly as a result of the radial 
segregation in the flowing layer, which keeps large particles near the surface.
Consequently, the return flow toward the equator deep in the flowing layer
consists of only small 
particles. Thus small particles reach the surface only near the 
equator where the core current emerges.
The consequence is a very sharp boundary at the surface between small and large 
particles for the SLS segregation pattern and a more diffuse 
boundary for the LSL case (see Fig.~\ref{trajbidicombi}). 
These differences are also evident in the experimental results in
Fig.~\ref{exproughsmooth} and \cite{ChenLueptow09}.

\subsection{Wall friction and roughness}

\begin{figure}[htbp]
\includegraphics[width=\linewidth]{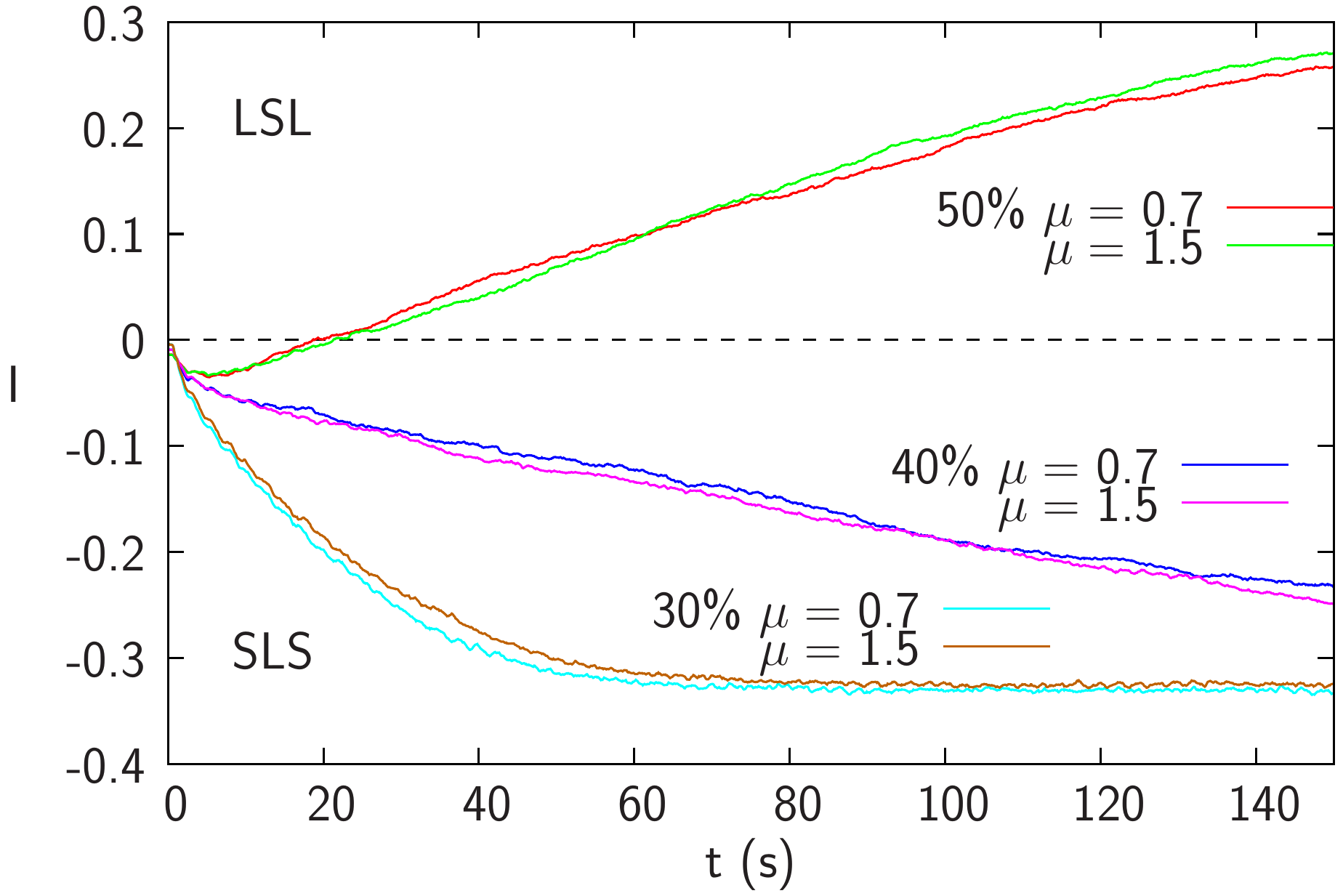}
\caption{(Color online) Evolution of the axial segregation index for 2~mm and 4~mm particles
rotated at 15~rpm for smooth walls but having wall
coefficients of friction of $\mu$=0.7 or 1.5 for fill fractions of 30\%, 
40\%, and 50\%.}
\label{coulomb}
\end{figure}

A question that naturally arises is if wall Coulomb friction could play a 
similar role to wall roughness in the evolution of segregation patterns based 
on the implicit assumption that a rough
wall should have a similar effect to a high coefficient of friction for a smooth
wall.  To consider this, we use a higher coefficient of friction for 
wall-particle interactions than for particle-particle interactions in 
simulations for smooth walls. As shown in
Fig.~\ref{coulomb}, the segregation evolution
for a wall coefficient of friction of 1.5 is nearly
identical to that for a wall coefficient of friction of 0.7, which is
the particle-particle coefficient of friction. Of course, increasing
the coefficient of friction too much results in a non-physical situation 
in which slip will not occur at all. Hence, it is difficult to explain the 
differences between smooth and rough walls based on a argument that Coulomb 
friction equivalent to wall roughness, at least within the constraints of 
the simulations. 

\begin{figure}[htbp]
\includegraphics[width=0.9\linewidth]{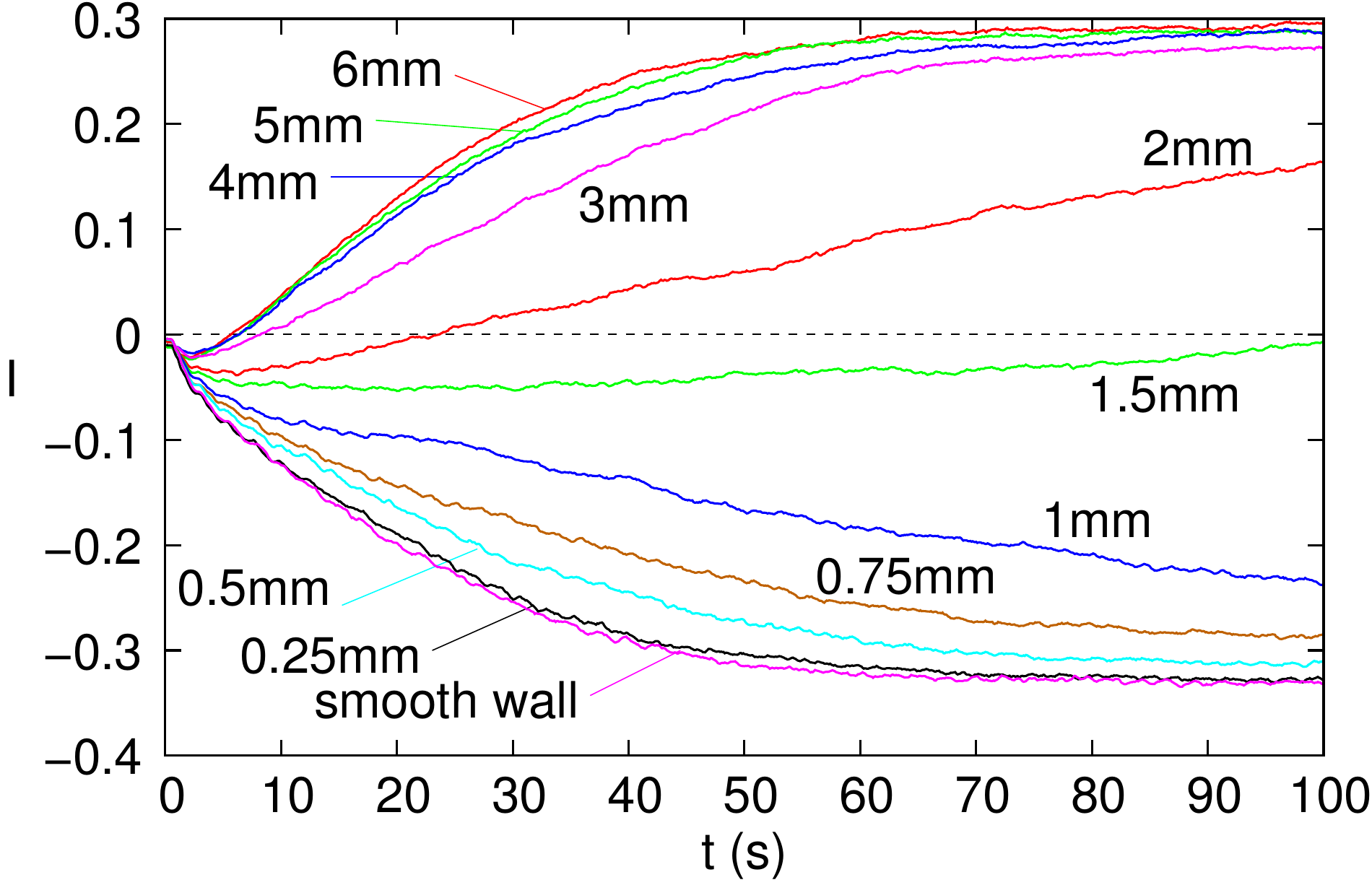}
\caption{(Color online) Comparison of the time evolution of the axial segregation index for 
2 and 4~mm particles at 30\% fill and different wall roughnesses.}
\label{segregvsrough5mm}
\end{figure}

\begin{figure}[htbp]
\includegraphics[width=0.9\linewidth]{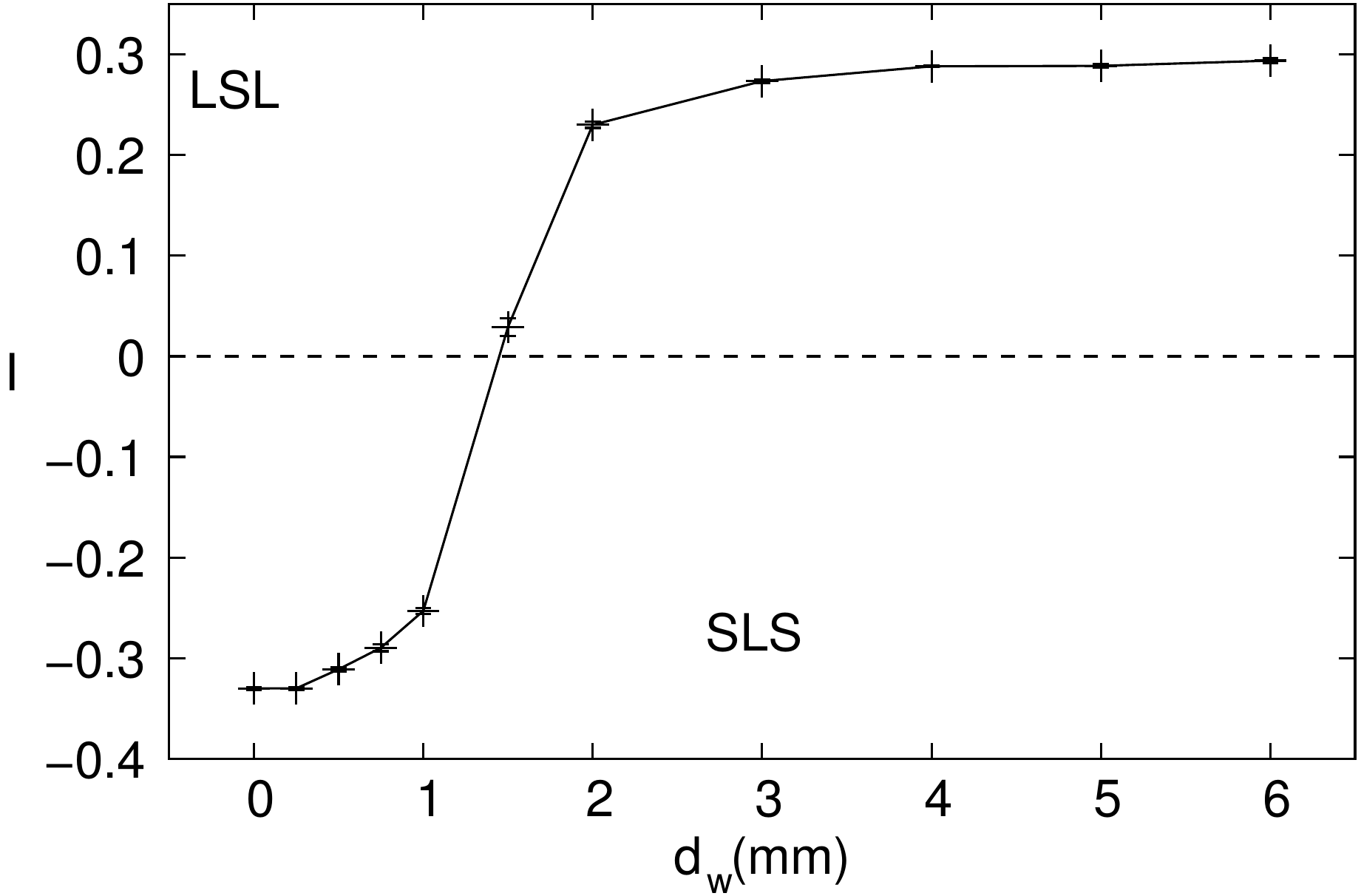}
\caption{Asymptotic value of the axial segregation index as function 
of the roughness of the wall for 2 and 4~mm particles at 30\% fill. Error bars 
(smaller than the symbols) are the standard deviation of the axial segregation index.}
\label{indexvsrough}
\end{figure}

Next, we consider the effect of variation of the roughness of the wall. 
The evolution of the axial segregation index for the different wall 
roughnesses and a fill level of 30\% is shown in Fig. \ref{segregvsrough5mm}.
As observed previously (Figs.~\ref{ios0-2mm} and \ref{charnin}) the segregation
 index increases with wall roughness. When the roughness is 
much larger than the size of the flowing particles, there is little influence
on the axial segregation index. This is likely because the smaller flowing
particles fill the gaps between the larger wall particles, which alters
particle trajectories \cite{DOrtonaThomas15}.
However, when the wall particles are smaller than the flowing particles, 
small changes can have significant impact. This is even more evident when
considering the asymptotic value of the segregation index, shown in 
Fig.~\ref{indexvsrough}. The greatest impact on the segregation index for 
a mixture of 2 and 4~mm particles occurs when the wall particle size is
between 1 and 2~mm.

It is also interesting to note that the perfectly smooth wall, modeled
as spherical smooth surface with infinite mass, behaves similarly to a 
rough wall of 0.25~mm particles, suggesting that modeling a wall as fixed
particles can be effective so long as the wall particles are much smaller
than the flowing particles.

\subsection{Size ratio}

To further consider the relative size effects of the tumbler and particles, 
we performed a limited number of
simulations with 1 and 2~mm particles in a 14~cm tumbler with both
smooth and 1~mm rough
walls.  Because many more particles are simulated in this situation than
with 2 and 4~mm particles, only low fill levels (30\%) and shorter runs were
\begin{figure}[htbp]
\includegraphics[width=0.99\linewidth]{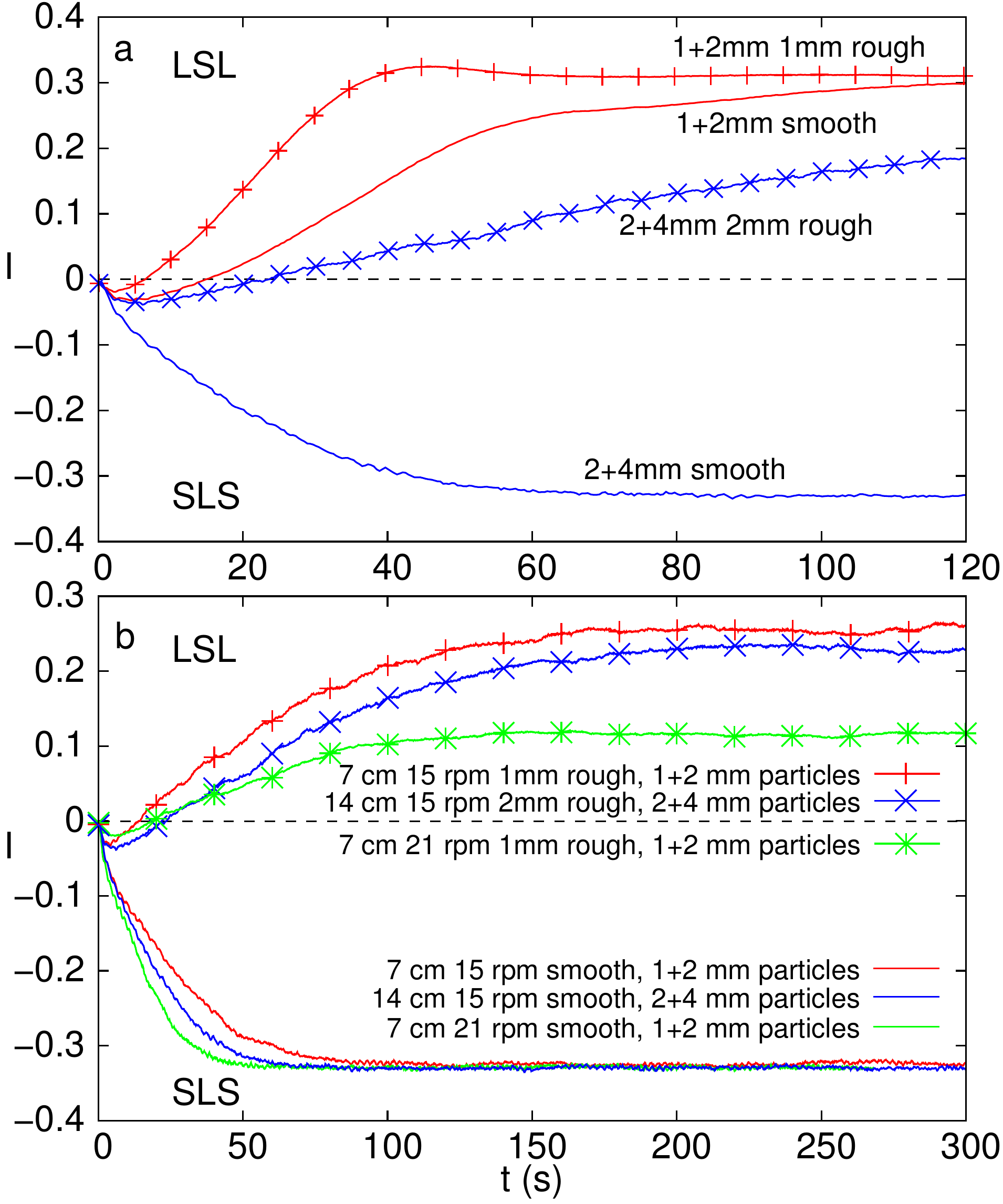}
\caption{(Color online) a) Comparison of the axial segregation index for a 
1+2~mm (red) and 2+4~mm (blue) system in a 14~cm tumbler filled at 30\%.
b) Two homothetic systems (at the same rotation speed or same Froude number):
1+2~mm particles 7~cm tumblers and 2+4~mm in 14~cm tumblers filled at 30\%.}
\label{chasize}
\end{figure}
feasible.  The results in Fig.~\ref{chasize}(a) demonstrate
that relative particle and tumbler sizes make a difference for both smooth
and rough walls.  The decrease in particle size induces an increase in 
the axial segregation index. For example, in the smooth wall case, 
the 1 and 2~mm particles form an LSL
pattern, while 2 and 4~mm particles form an SLS pattern, consistent with
experimental results \cite{ChenLueptow09}.  
This is because particle trajectory curvature increases with particles 
size, but net axial drift remains almost independent of particles size 
\cite{DOrtonaThomas15}, and large curvature trajectories
favor SLS segregation. With trajectory curvature being smaller for small 
particles, a 1+2~mm particle system will adopt the LSL segregation pattern
more easily than a 2+4~mm system, as is observed 
experimentally \cite{ChenLueptow08} and shown in Fig.~\ref{chasize}(a).

Consider now a 7~cm tumbler (30\% full) in which the entire system  
(particle sizes, wall roughness, and tumbler diameter) is scaled to half the 
size of the previous simulations, compared to a 30\% full 2+4~mm system at
the same rotation speed.
The differences between the large and small
systems are relatively small as shown in Fig.~\ref{chasize}(b). 
Results for the same Froude number, $Fr=\omega^2 R/g$, for the 7~cm
tumbler at 21~rpm and the 14~cm tumbler 15~rpm are also shown in
Fig.~\ref{chasize}(b).
For now, we simply note that conserving the rotation speed while reducing 
the system size results in similar axial segregation index curves while 
conserving the Froude Number results
in different degrees of axial segregation for different system sizes.
Changing the drum size suggests that the key mechanism is 
probably the differences in the trajectories for large and small 
particles.  

\section{Conclusions}

Bidisperse particle segregation in a spherical tumbler provides an ideal test 
for evaluating the impact of wall boundary
conditions because of its sensitivity to the wall roughness and easily 
visualized results. It
is clear that surface boundary conditions have a strong influence 
on the flow and subsequent segregation patterns.
The bands form due to a combination of curved particle
trajectories and the axial drift in the flowing layer.  While the roughness
of the walls determines the curvature and drift, these quantities along
with radial size segregation determine the nature of the
axial segregation pattern.  For adequate fill levels, smooth walls result in 
more
curved trajectories with little drift and consequently SLS patterns;
rough walls result in less curved trajectories with more drift and
consequently LSL patterns.  At large fill levels, the axial 
LSL patterns always occur regardless of wall roughness. At lower
fill levels, axial SLS patterns are more likely to occur.

Many questions remain including that of why the curvature and drift are
so dependent upon the size of the system and rotation speed. 
Nevertheless,  the non-locality of granular flow (wall roughness modifies the
trajectories and band formation far from the wall) is evident, as
is the case in many other situations for granular pattern formation.


\begin{thebibliography}{99}
\bibitem{MoakherShinbrot00}M. Moakher, T. Shinbrot, and F. J. Muzzio, Powder Technol. {\bf 109}, 58 (2000).
\bibitem{ChenLueptow11}P. Chen, J. M. Ottino, and R. M. Lueptow, New J. Phys. {\bf 13}, 055021 (2011).
\bibitem{TaberletLosert04}N. Taberlet, W. Losert, and P. Richard, Europhys. Lett. {\bf 68}, 522 (2004).
\bibitem{Rapaport02}D. C. Rapaport, Phys. Rev. E {\bf 65}, 061306 (2002).
\bibitem{TaberletNewey06}N. Taberlet, M. Newey, P. Richard, and W. Losert, J. Stat. Mech. P07013 (2006).
\bibitem{DaCruzEmam05}F. da Cruz, S. Emam, M. Prochnow, J.-N. Roux, and F. Chevoir, Phys. Rev. E {\bf 72}, 021309 (2005).
\bibitem{McCarthyOttino98}  J. J. McCarthy, and J. M. Ottino, Powder Technol. {\bf 97}, 91 (1998).
\bibitem{PoschelBuchholtz95} T. P\"oschel and V. Buchholtz, Chaos, Solitons and Fract. {\bf 5}, 1901 (1995).
\bibitem{JuarezChen11} G. Juarez, P. Chen, and R. M. Lueptow, New J. Phys. {\bf 13}, 053055 (2011).
\bibitem{BertrandLeclaire05} F. Bertrand, L. A. Leclaire, and G. Levecque, Chem. Eng. Sci. {\bf 60}, 2517 (2005).
\bibitem{DOrtonaThomas15}U. D'Ortona, N. Thomas, Z. Zaman, and R. M. Lueptow, Influence of Rough and Smooth Walls on Macroscale Flows in Tumblers, submitted to Phys. Rev. E
\bibitem{TaberletRichard03}N. Taberlet, P. Richard, A. Valance, W. Losert, J. M. Pasini, J. T. Jenkins, and R. Delannay, Phys. Rev. Lett. {\bf 91}, 264301 (2003).
\bibitem{MeierLueptow07}S. W. Meier, R. M. Lueptow, and J. M. Ottino, Adv. Phys. {\bf 56}, 757 (2007).
\bibitem{duPontGondret03}S. Courrech du Pont, P. Gondret, B. Perrin, and M. Rabaud, Europhys. Lett. {\bf 61}, 492 (2003).
\bibitem{PignatelLueptow12}F. Pignatel, C. Asselin, L. Krieger, I. C. Christov, J. M. Ottino, and R. M. Lueptow, Phys. Rev. E {\bf 86}, 011304 (2012).
\bibitem{GilchristOttino03}J. F. Gilchrist and J. M. Ottino, Phys. Rev. E {\bf 68}, 061303 (2003).
\bibitem{ChenLueptow09}P. Chen, B. J. Lochman, J. M. Ottino, and R. M. Lueptow, Phys. Rev. Lett., {\bf 102}, 148001 (2009).
\bibitem{JuarezLueptow08}G. Juarez, J. M. Ottino, and R. M. Lueptow, Phys. Rev. E {\bf 78}, 031306 (2008).
\bibitem{Oyama39}Y. Oyama, Bull. Inst. Phys. Chem. Res. (Tokyo) {\bf 18}, 600 (1939).
\bibitem{DonaldRoseman62}M. B. Donald and B. Roseman, Br. Chem. Eng. {\bf 7}, 749 (1962).
\bibitem{Nakagawa94}M. Nakagawa, Chem. Eng. Sci. {\bf 49}, 2540 (1994).
\bibitem{ZikLevine94}O. Zik, D. Levine, S. G. Lipson, S. Shtrikman, and J. Stavans, Phys. Rev. Lett. {\bf 73}, 644 (1994).
\bibitem{HillKakalios94}K. M. Hill and J. Kakalios, Phys. Rev. E {\bf 49}, R3610 (1994).
\bibitem{HillKakalios95}K. M. Hill and J. Kakalios, Phys. Rev. E {\bf 52}, 4393 (1995).
\bibitem{JainLueptow01}N. Jain, D. V. Khakhar, R. M. Lueptow, and J. M. Ottino, Phys. Rev. Lett. {\bf 86}, 3771 (2001).
\bibitem{FiedorOttino03}S. J. Fiedor and J. M. Ottino, Phys. Rev. Lett. {\bf 91}, 244301 (2003).
\bibitem{ChenLueptow10}P. Chen, J. M. Ottino, and R. M. Lueptow, Phys. Rev. Lett. {\bf 104}, 188002 (2010).
\bibitem{ChenLueptow08}P. Chen, J. M. Ottino, and R. M. Lueptow, Phys. Rev. E {\bf78}, 021303 (2008).
\bibitem{SchaferDippel96}J. Sch\"afer, S. Dippel, and D. E. Wolf, J. Phys. I (France) {\bf 6}, 5 (1996).
\bibitem{Ristow00}G. H. Ristow, Pattern Formation in Granular Materials (Springer-Verlag, Berlin, 2000).
\bibitem{CundallStrack79} P. A. Cundall and O. D. L. Strack, Geotechnique {\bf 29}, 47 (1979).
\bibitem{AllenTildesley02}M. P. Allen and D. J. Tildesley, Computer Simulation of Liquids (Oxford University Press, New York, 2002).
\bibitem{DrakeShreve86}T. G. Drake and R. L. Shreve, J. Rheol. {\bf 30}, 981 (1986).
\bibitem{FoersterLouge94}S. F. Foerster, M. Y. Louge, H. Chang, and K. Allia, Phys. Fluids {\bf 6}, 1108 (1994).
\bibitem{ZamanDOrtona13}Z. Zaman, U. D'Ortona, P. B. Umbanhowar, J. M. Ottino, and R. M. Lueptow, Phys. Rev. E {\bf 88}, 012208 (2013).
\bibitem{Campbell02}C. S. Campbell, J. Fluid Mech. {\bf 465}, 261 (2002).
\bibitem{SilbertGrest07}L. E. Silbert, G. S. Grest, R. Brewster, and A. J. Levine, Phys. Rev. Lett. {\bf 99}, 068002 (2007).
\end{thebibliography}
\end{document}